\documentclass{article}

\usepackage{amsmath,amsfonts}
\usepackage{graphicx}
\usepackage{algorithm,algorithmic}
\usepackage{hyperref}

\newcommand{\R}{\mathbb{R}}
\newcommand{\N}{\mathbb{N}}

\newtheorem{theorem}{Theorem}

\title{Frequent Knot Discovery}
\author{Floris Geerts\\
Laboratory for Foundations of Computer Science\\
School of Informatics\\
University of Edinburgh, UK\\
fgeerts@inf.ed.ac.uk}

\begin{document}

\maketitle
\begin{abstract}
We explore the possibility of applying the framework of frequent
pattern mining to a class of continuous objects appearing in
nature, namely knots.
 We introduce the frequent knot mining problem and
present a solution. The key observation is that a database
consisting of knots can be transformed into a transactional
database. This observation is based on the Prime Decomposition
Theorem of knots.
\end{abstract}

\section{Introduction}

\begin{figure}[b]
\centering
\includegraphics[height=3cm]{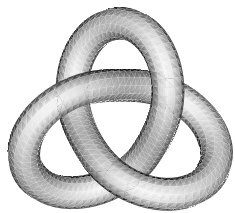}
\caption{Trefoil knot.}\label{fig:trefoil}
\end{figure}

Many algorithms have recently been developed for mining frequent
patterns. Traditionally, these patterns consist of subsets of
attributes in a relational database \cite{Agrawal}. Recently,
other patterns have been mined, such as trees~\cite{Zaki} and
graphs~\cite{Kuramochi-01,Kuramochi-02,Yan,Inokuchi}. However,
most objects appearing in nature lack the discrete character of
graph and trees. In this paper we explore the possibility of
applying the framework of frequent pattern mining to a class of
continuous objects appearing in nature, namely knots. A knot can
be thought of as a piece of rope (where the rope has zero
thicknes) which forms a loop in three-dimensional Euclidean space
$\R^3$. Figure~\ref{fig:trefoil} shows an example of a knot known
as the Trefoil knot.

The history of knots dates back to the late 1800's when Lord
Kelvin suggested that atoms where knots in an invisible and
frictionless fluid. Since then, theoretical properties of knots
are extensively studied in mathematics~\cite{Burde}. In physics,
knot invariants (e.g., the Jones polynomial) are used in
statistical physics ~\cite{Kauf} and knots also appear in the
context of quantum gravity~\cite{Baez}. Recently, knots showed up
as building blocks for future quantum computers~\cite{Lomonaco}.

In biology, knots are used to characterize topoisomerase
enzymes~\cite{Dewitt} and in polymer science, physical properties
of long ring polymers, such as DNA, gels and rubbers are related
to properties of knots~\cite{Gennes,Doi,Whittington}. It is shown
that knots are present in such polymers with probability one when
the polymers are long enough~\cite{Sumn}.  There is also much
interest in developing artificial knotted biopolymers as building
blocks for DNA-based computing~\cite{Seeman}.

The study of knotted polymers is done both by using experimentally
obtained knots and by using knots obtained by numerical approaches
based on self-avoiding random-walk simulations. Examples of
questions one would like to answer in these studies are what is
the probability of having a certain knot  in polymers of a certain
length~\cite{Shimamura}, and whether the knots appear tight or
loose in the knotted polymers~\cite{Katritch}.

In this article we consider the {\em frequent knot mining problem}
which can be stated as follows: Given a collection of knots, find
all subknots which appear  frequently in this collection.

We believe that finding frequent subknots in a large collection of
real or simulated knotted polymers, will contribute to a deeper
understanding of the statistical properties of knotted polymers in
$\R^3$. This article reports a first attempt for solving the
frequent knot mining problem.

The solution presented in this article consists of three steps:
\begin{enumerate}
\item Encoding of knots in transactions; \item Mining these
transactions; and finally, \item Decoding of the frequent itemsets
into knots.
\end{enumerate}

The article is organized as follows: In Section 2, definitions are
given and the frequent knot mining problem is stated formally. The
encoding (decoding) of knots (transactions) into transactions
(knots) is described in Section 3 . In Section 4, we present the
KnotMiner algorithm for mining frequent knots. Finally,
conclusions are drawn in Section 5.

\section{Preliminaries}

A {\em knot} $K$ can be thought of as a piece of rope (where the
rope has zero thickness) which forms a loop in three-dimensional
Euclidean space $\R^3$. Two knots $K$ and $K'$ are equivalent, or
in symbols $K\equiv K'$, if they can be transformed into each
other without cutting and pasting the ropes. We will only consider
so-called tame knots. These are knots which are equivalent to
piecewise linear knots, i.e., knots consisting of a finite number
of straight lines. A knot is trivial if it is equivalent to a rope
which forms a circle in a plane in $\R^3$.

A knot can be finitely represented by a {\em knot diagram}. The
knot diagram of a knot $K$ is a connected undirected planar graph,
which correspond to a (generic) projection of $K$ onto a plane.
Vertices in a knot diagram correspond to places where the
projection of the knot intersects, and each edge adjacent to a
vertex is labelled as an undercrossing or overcrossing, whichever
is the case. Given a knot diagram consisting of $n$ vertices one
can find in time polynomial in $n$ a piecewise linear knot such
that its $z$-projection gives the original diagram~\cite{Hass}.
Two knot diagrams  can be transformed into each other using the
so-called  Reidemeister moves if and only if they represent
equivalent knots~\cite{Burde}.

\begin{figure}
\centering
\includegraphics[height=3cm]{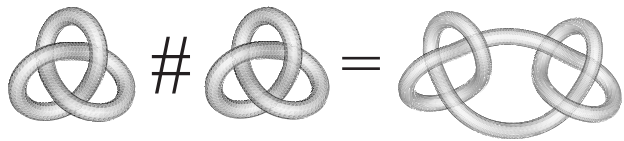}
\caption{Sum of two knots.}\label{fig:sum}
\end{figure}

The {\em connected sum} of two knots $K_1$ and $K_2$ is formed by
removing a small piece of rope from both knots and then connecting
the four endpoints by two new pieces of rope in such a way that no
new crossings are introduced, the result being a a single knot,
which is denoted by $K=K_1\# K_2$. This operation is illustrated
in Figure~\ref{fig:sum}. The connected sum $K_1\#K_2$ is
equivalent to $K_2\#K_1$ and $(K_1\#K_2)\#K_3$ is equivalent to
$K_1\#(K_2\#K_3)$. The connected sum of a knot $K$ and the trivial
knot is equivalent to $K$~\cite{Burde}.

A knot is called {\em prime} if for any decomposition as a
connected sum, one of the factors is the trivial knot. There are
infinitely many  prime knots.

\begin{theorem}[Prime decomposition Theorem~\cite{Burde,Schubert}]
Every knot $K$ can be decomposed as a connected sum of nontrivial
prime knots. If $K\equiv K_1\# K_2\#\allowbreak \cdots
\allowbreak\# K_m$ and $K\equiv L_1\# L_2\# \cdots \# L_n$, where
$K_i$ and $L_i$ are nontrivial prime knots, then $m=n$, and after
reordering each $K_i$ is equivalent to $L_i$.
\end{theorem}

This theorem motivates the following definition. Let
$$
K= K_1\# K_2\#\cdots\# K_p
$$
and let
$$
L=L_1\# L_2\# \cdots\# L_q.
$$
Then, $K$ is a {\em subknot} of $L$, or $K\preceq L$, if for any
$i=1,\ldots,p$ we have that
$$
\left|\{j\mid K_j\equiv K_i\}\right| \leq \left|\{j\mid L_j\equiv
K_i\}\right|.
$$

A {\em knot datatabase} ${\mathcal D}$ is a finite collection of
knots. The {\em support} of a knot $K$ in ${\mathcal D}$ is
defined as
$$
{\sf supp}(K)=| \{ L \in {\mathcal D}\mid K\preceq L\}|.
$$

The {\em frequent knot mining problem} can be stated as follows:
Given a knot database ${\mathcal D}$ and a threshold value
$\sigma\in\N$, find all knots $K$ in ${\mathcal D}$ such that
${\sf supp}(K)>\sigma$.

For completeness, we also state the frequent itemset mining
problem. A {\em transaction database} ${\mathcal T}$ is a finite
collection of $k$-tuples in  $\N^k$. The {\em support} of an
itemset $I$ in ${\mathcal T}$ is defined as $$ {\sf supp}(I)=|\{
J\in{\mathcal T}\mid \forall\ell :(I)_\ell\leq (J)_\ell\}|, $$
where $(I)_\ell$ (resp. $(J)_\ell$) denotes the $\ell$th component
of $I$ (resp. $J$). The {\em frequent itemset mining problem} is
then: Given a transaction database ${\mathcal T}$ and a threshold
value $\sigma\in\N$, find all itemsets $I$ in ${\mathcal T}$ such
that ${\sf supp}(I)>\sigma$.

\section{From Knot Databases to Transaction Databases}
In this section we show how to transform a knot database
${\mathcal D}$ into a transactional database. We assume that the
knots in ${\mathcal D}$ are represented by knot diagrams.

We start by  computing for each knot $K$ in ${\mathcal D}$ its
prime decomposition. Schubert~\cite{Schubert} gives an algorithm
computing this decomposition. The running time is at worst
exponential in the number of vertices in the knot diagram. In this
way, we obtain a set ${\sf primes}({\mathcal D})$ consisting of
knot diagrams for all prime knots occurring in ${\mathcal D}$. Two
different knot diagrams in ${\sf primes}({\mathcal D})$ can
represent the same prime knot, so we have to eliminate duplicates.
There exists an algorithm for testing whether two knot diagrams
represent equivalent knots~\cite{Hemion,Waldhausen}. However, at
present, the complexity of this algorithm is not known. From here
on, we assume that ${\sf primes}({\mathcal D})$ does not contain
duplicates and order it arbitrarily.

We now define a mapping, denoted by  $ {\sf encode} $, from knots
in a knot database to elements in a transaction database. Let
${\mathcal D}$ be a knot database, and $K$ a knot in ${\mathcal
D}$. Then,
$${\sf encode}(K)=(n_1,\ldots,n_p),$$ with $p=|{\sf primes}({\mathcal
D})|$ and $n_i$ is the number of times the prime knot
corresponding to the  $i$th knot diagram in ${\sf
primes}({\mathcal D})$ appears in the prime decomposition of $K$.
Clearly, ${\sf encode}({\mathcal D})$ is a transaction database
consisting of $|{\sf primes}({\mathcal D})|$ attributes.

Given a set of knots ${\mathcal K}=\{K_1,\ldots,K_p\}$, we now
define the mapping, denoted by ${\sf decode}$, from  itemsets of a
transaction database ${\mathcal T}\subset\N^p$ to knots in $\R^3$.
Let $t\in{\mathcal T}$ and let
$(t)_{i_1,\ldots,i_k}=(m_1,\ldots,m_k)$ be a $k$-itemset.
 Then,
$$
{\sf decode}(m_1,\ldots,m_k)=\underbrace{K_{i_1}\#\cdots\#
K_{i_1}}_{m_1 \text{ times }}\#\cdots \#
\underbrace{K_{i_k}\#\cdots\#K_{i_k}}_{m_k\text{ times}}.
$$
For any knot $K$, we have that $K\equiv{\sf decode}({\sf
encode}(K))$.

\section{Algorithm}

We now present The KnotMiner algorithm for computing the frequent
knots in a knot database.

\begin{algorithm}
\caption{KnotMiner}
\begin{algorithmic}[1]
\REQUIRE knot database $\mathcal{D}, \sigma$ \ENSURE All knots $K$
such that ${\sf supp}(K)>\sigma$. \STATE Compute ${\mathcal
T}:={\sf encode}({\mathcal D})$ \STATE Compute the set ${\mathcal
F}$ of frequent itemsets in ${\mathcal T}$ \STATE Output ${\sf
decode}({\mathcal F})$.
\end{algorithmic}
\end{algorithm}
The first and last step in KnotMiner are already fully explained
in Section III. For the second step one can either transform
${\mathcal T}$ into a binary transaction database and use a
standard mining algorithm like Apriori~\cite{Agrawal},
Eclat~\cite{Zaki2} or FP-growth~\cite{Han}. Alternatively one can
mine ${\mathcal T}$ directly using  algorithms presented
in~\cite{Srikant} and~\cite{Possas}. The following result is
immediate.
\begin{theorem}
The KnotMiner algorithm works correctly.
\end{theorem}
\section{Concluding Remarks and Future Work}
In this article we introduced the frequent knot mining problem and proposed the KnotMiner
algorithm to solve it. Currently, there exists no implementation of KnotMiner. This is mainly
due to the complex algorithms needed for the encoding of a knot database into a transactional databases.

However, recent research indicates that the knot decomposition of knotted polymers
can be obtained by ``Coulomb decomposition'', which is a technique where polymers
are brought into an equilibrium state using Coulomb interactions~\cite{Dommer}. We
hope to apply this technique on simulated knotted polymers and hence obtain an
implementation of KnotMiner, specifically aimed  for mining knotted polymer databases.


\end{document}